\documentclass[twocolumn]{aastex63}

\usepackage{lmodern}
\usepackage{latexsym}
\usepackage{amsmath}
\usepackage{amssymb}
\usepackage{amsbsy}
\usepackage{amsthm}
\usepackage{amsfonts}
\usepackage{mathrsfs}
\usepackage{bm}
\usepackage{sansmath}
\usepackage{relsize}
\usepackage{caption2}
\usepackage{graphicx}
\usepackage[utf8]{inputenc} 
\usepackage[T1]{fontenc}
\usepackage{epstopdf}
\usepackage{lineno}

\received{2021 April 22}
\accepted{2021 May 31}
\shorttitle{Self-similar solution of hot accretion flow}
\shortauthors{Zeraatgari et al.}

\begin{document}

\title{Self-similar solution of hot accretion flow: the role of kinematic viscosity coefficient}

\correspondingauthor{Liquan Mei}
\email{lqmei@mail.xjtu.edu.cn}

\author[0000-0003-3345-727X]{Fatemeh Zahra Zeraatgari}
\affiliation{School of Mathematics and Statistics, Xi'an Jiaotong University, Xi'an, Shaanxi 710049, PR China}

\author[0000-0003-3468-8803]{Liquan Mei}
\affiliation{School of Mathematics and Statistics, Xi'an Jiaotong University, Xi'an, Shaanxi 710049, PR China}

\author[0000-0002-4601-7073]{Amin Mosallanezhad}
\affiliation{School of Mathematics and Statistics, Xi'an Jiaotong University, Xi'an, Shaanxi 710049, PR China}



\begin{abstract}

We investigate the dependency of the inflow-wind structure of the 
hot accretion flow on the kinematic viscosity coefficient. 
In this regard, we propose a model for the kinematic viscosity coefficient to mimic 
the behavior of the magnetorotational instability and would be maximal 
at the rotation axis. Then, we compare our model with two other prescriptions 
from numerical simulations of the accretion flow. We solve
two-dimensional hydrodynamic equations of hot accretion flows in the presence 
of the thermal conduction. The self-similar approach is also adopted 
in the radial direction. We calculate the properties of the inflow and 
the wind such as velocity, density, angular momentum for three models 
of the kinematic viscosity prescription. On inspection, we find that 
in the model we suggested wind is less efficient than that in two other models
to extract the angular momentum outward where the self-similar solutions are applied. 
The solutions obtained in this paper might be applicable to the  
hydrodynamical numerical simulations of the hot accretion flow.

\end{abstract}



\section{Introduction} \label{sec:intro}

Based on the temperature of the flow, black hole accretion disks 
can be divided to two broad modes: hot and cold. The standard 
thin disk as well as slim disk models, with relatively high mass 
accretion rates, are belonged to the cold mode (\citealt{Shakura 
and Sunyaev 1973}; \citealt{Abramowicz et al. 1988}), while the 
radiative inefficient accretion flow with low mass accretion rate, 
below roughly $ 2 \% \dot{M}_{\mathrm{Edd}} $, is belonged to 
the hot mode ($ \dot{M}_{\mathrm{Edd}}  = 10 L_{\mathrm{Edd}}  / c^{2} $ 
is called the Eddington accretion rate with $ L_{\mathrm{Edd}} $ 
and $ c $ being the Eddington luminosity and speed of light, 
respectively.) (\citealt{Narayan and Yi 1994}). 

Wind as an essential constituent of accretion flows can impact 
the dynamics and structure of that. In addition, the feedback of 
the wind on the surrounding medium can also play a major role 
in formation and evolution of the galaxies (\citealt{Fabian 2012}; 
\citealt{Kormendy and Ho 2013}; \citealt{Naab and Ostriker 2017}).
It has been generally acknowledged that winds are prevailing in 
hot accretion flows. The fully-ionized winds from hot accretion 
flows have been confirmed by the wind observations from 
low-luminous active galactic nuclei (LLAGNs) as well as radio 
galaxies (\citealt{Tombesi et al. 2010, Tombesi et al. 2014}; 
\citealt{Crenshaw and Kraemer 2012}; \citealt{Cheung et al. 2016}), 
the supermassive black hole in the Galactic center (Sgr A*; 
\citealt{Wang et al. 2013}; \citealt{Ma et al. 2019}), and black 
hole X-ray binaries in a hard state in recent years 
(\citealt{Homan et al. 2016}; \citealt{Munoz-Darias et al. 2019}).

Three different wind-lunching mechanics have been proposed, 
namely thermally driven (\citealt{Begelman et al. 1983}; \citealt{Font et al. 2004}; 
\citealt{Luketic et al. 2010}; \citealt{Waters and Proga 2012}), 
magnetically driven (\citealt{Blandford and Payne 1982}; 
\citealt{Lynden-Bell 1996, Lynden-Bell 2003}), and radiation 
driven wind (\citealt{Murray et al. 1995}; \citealt{Proga et al. 2000}; 
\citealt{Proga and Kallman 2004}; \citealt{Nomura and Ohsuga 2017}).
In terms of hot accretion flow the first two mechanisms can play 
a role for producing wind since the radiation is negligible.
Therefore, hot accretion flows are more feasible to be simulated. 
To show the existence of wind in hot accretion flow, a large number 
of global hydrodynamical (HD) and magnetohydrodynamical (MHD) 
numerical simulations have been done ( e.g., \citealt{Stone et al. 1999}; 
\citealt{Igumenshchev and Abramowicz 1999, Igumenshchev and 
Abramowicz 2000}; \citealt{Stone and Pringle 2001}; \citealt{Hawley et al. 2001}; 
\citealt{Machida et al. 2001}; \citealt{Hawley and Balbus 2002}; 
\citealt{Igumenshchev et al. 2003}; \citealt{Pen et al. 2003}; 
\citealt{De Villiers et al. 2003, De Villiers et al. 2005}; 
\citealt{Yuan and Bu 2010}; \citealt{Pang et al. 2011}; 
\citealt{McKinney et al. 2012}; \citealt{Narayan et al. 2012}; 
\citealt{Li et al. 2013}; \citealt{Yuan et al. 2012a,Yuan et al. 2012b}; 
\citealt{Yuan et al. 2015}; \citealt{Inayoshi et al. 2018, Inayoshi et al. 2019}).

The pioneer analytical studies of hot accretion flow such as 
\citealt{Narayan and Yi 1994, Narayan and Yi 1995}; 
\citealt{Blandford and Begelman 1999, Blandford and Begelman 2004};
\citealt{Xu and Chen 1997} and \citealt{Xue and Wang 2005} have 
also claimed that the strong wind must exist. To deep understanding 
the physical properties of inflow and wind from accretion flow, 
analytical studies are powerful tools. This is mainly because 
it is very time-consuming and difficult to run global numerical 
simulations of accretion flow with different physics including 
dissipation and magnetic field terms. While it is much easier to 
probe the dependency of the results to the input parameters in 
analytical approach. Based on radial self-similar approximation, 
the existence of wind have been investigated in more details 
by solving HD and MHD equations of hot accretion flow 
(e.g., \citealt{Bu et al. 2009}; \citealt{Jiao and Wu 2011}, 
\citealt{Mosallanezhad et al. 2014}; \citealt{Samadi et al. 2017}; 
\citealt{Bu and Mosallanezhad 2018}; \citealt{Kumar and Gu 2018}; 
\citealt{Zeraatgari et al. 2020}). Due to the technical difficulties 
such as the existing of singularity near the rotation axis in these 
studies the physical boundary conditions set at only the equatorial 
plane and the equations have been integrated from this boundary. 
Then, the integration stop where the gas pressure or density 
becomes zero for HD case (see e.g. \citealt{Jiao and Wu 2011}) 
or where the sign of the radial or toroidal components of the 
magnetic field changes in MHD case (see e.g., \citealt{Mosallanezhad et al. 2016}). 
To get reliable physical solutions, similar to \citealt{Narayan and Yi 1995} 
and \citealt{Xu and Chen 1997}, a second boundary should be considered 
at the rotation axis and the equations would be solved using 
two-point boundary value problem technique.

In \citealt{Zeraatgari et al. 2020} and \citealt{Mosallanezhad et al. 2021}, 
we utilized such a technique to solve the radiation hydrodynamic (RHD) 
equations of supercritical as well as HD equations of hot accretion flows, 
respectively. More precisely, in \citealt{Mosallanezhad et al. 2021}, 
we solved the HD equations of the accretion flow with thermal conduction 
using relaxation method. We explained the origin 
of the discrepancy between previous self-similar solutions of 
hot accretion flow in the presence of the wind such as \citealt{Xu and Chen 1997} 
and \citealt{Xue and Wang 2005} by extensively study the energy 
equation. We Showed that in terms of hot accretion flow, thermal 
conduction would be an essential term for investigating the inflow-wind 
structure of the flow. Another interesting result was that the hot 
accretion flow is convectively stable in the presence of the thermal conduction. 

It is now widely believed that in a real accretion flow the angular momentum 
is transferred with the MHD turbulence driven by the magnetorotational instability, 
MRI (\citealt{Balbus and Hawley 1991, Balbus and Hawley 1998}). 
Consequently, the MHD equations should be solved rather than HD ones. 
In \citealt{Mosallanezhad et al. 2021}, instead of magnetic field, we 
considered the anomalous shear stress to mimic the magnetic stress. 
For the kinematic viscosity coefficient, following \citealt{Narayan 
and Yi 1995}, we only adopted $ \alpha $-prescription as $ \nu = \alpha c_{s}^{2} 
/ \Omega_\mathrm{K} $ where $ c_{s} $ and $ \Omega_\mathrm{K} $ 
are sound speed and Keplerian angular velocity, respectively 
(\citealt{Shakura and Sunyaev 1973}). There are also some MHD 
numerical simulations which show the dependency of the 
magnetic stress on the vertical and radial structure of the disk 
( e.g., \citealt{Bai and Stone 2013}; \citealt{Penna et al. 2013}).
Therefore, following numerical simulations of hot accretion flow, 
we will adopt several different viscosity prescriptions and solve 
the vertical structure of the disk by using relaxation method.
It would be worthwhile to investigate the dependency of the 
physical quantities of the hot accretion flow as well as the wind 
properties such as poloidal velocity or Bernoulli parameter to 
the definition of the kinematic viscosity coefficient. 

The remainder of the manuscript is organized as follows. In section 
\ref{sec:basic_equations}, the basic equations, physical assumptions, 
self-similar solutions, and the boundary conditions will be introduced. 
The detailed explanations of numerical results to the definition of the 
viscosity coefficient will be presented in section \ref{sec:results}. Finally, 
in Section \ref{sec:summary_discussion} we will provide the summary 
and discussion. 

\section{Basic Equations and Assumptions} \label{sec:basic_equations}
The basic HD equations of the hot accretion flow can be described as

\begin{equation} \label{eq:continuity}
  \frac{ \mathrm{d} \rho}{\mathrm{d} t} +\rho \bm{\nabla} \cdot \bm{v} = 0,
\end{equation}

\begin{equation}\label{eq:momentum}
\rho  \frac{\mathrm{d} \bm{v}}{\mathrm{d} t} = - \rho \bm{\nabla} \psi  - 
\bm{\nabla} p_\mathrm{gas} +\bm{\nabla} \cdot \bm{\mathrm{\sigma}}, 
\end{equation}

\begin{equation}\label{eq:energy}
	\rho \frac{\mathrm{d} e}{\mathrm{d} t} - \frac{p_\mathrm{gas}}{\rho} \frac{\mathrm{d} \rho}{\mathrm{d} t}  =
	 \, \bm{\nabla} \bm{v} : \bm{\sigma} - \bm{\nabla} \cdot \bm{F}_\mathrm{c}.
\end{equation}

In the above equations, $ \rho $ is the mass density, $ \bm{v} $ is the velocity, $ \psi (= - GM / r) $ 
is the Newtonian potential (where $ r $ is the distance from the central black hole, 
$ M $ is the black hole mass and $ G $ is the gravitational constant), 
$ p_\mathrm{gas} $ is the gas pressure, $\bm{\sigma} $ is the viscous stress tensor, $ e $ is the gas internal energy, 
and $ \bm{F}_\mathrm{c} $ is the thermal conduction.
Here, $ \mathrm{d}/\mathrm{d} t \equiv \partial / \partial t +  \bm{v} \cdot \nabla $ represents
the Lagrangian or comoving derivative. The equation of state of the ideal gas is 
considered as $ p_\mathrm{gas} =  \left( \gamma - 1 \right) \rho e $, with $ \gamma = 5/3 $ 
being the adiabatic index. In purely HD limit, like our case, $ \bm{F}_\mathrm{c} $ is defined as

\begin{equation} \label{F_conduction}
	\bm{F}_\mathrm{c} = - \lambda \nabla T,
\end{equation}
where $ \lambda $ is the thermal diffusivity and $ T $ is the gas temperature.
We use spherical coordinates $ (r, \theta, \phi) $ to solve the above set 
of equations. In most previous semi-analytical studies on 
hot accretion flow, they assumed that the system is in steady state and axisymmetric,
i.e., $ \partial/\partial \phi = \partial/\partial t  = 0 $. These 
assumptions imply that all physical quantities are independent of the 
azimuthal angle, $ \phi $, and the time, $ t $. Following numerical simulations of
hot accretion flow (e.g., \citealt{Stone et al. 1999}; \citealt{Yuan et al. 2012a}), we consider 
the following components of the viscous stress tensor:

\begin{equation} \label{sigma_rt}
	\sigma_{r \theta} = \rho \nu  \left[ r \frac{\partial}{\partial r} \left( \frac{v_{\theta}}{r} \right) + \frac{1}{r} \frac{\partial v_{r}}{\partial \theta} 
	 \right], 
\end{equation}

\begin{equation} \label{sigma_rp}
	\sigma_{r \phi} = \rho \nu \left[ r \frac{\partial}{\partial r} \left(\frac{v_{\phi}}{r} \right)  \right],
\end{equation}

\begin{equation} \label{sigma_tp}
	\sigma_{\theta \phi} = \rho \nu \left[ \frac{\sin \theta}{r}  \frac{\partial}{\partial \theta}
	\left( \frac{v_{\phi}}{\sin \theta} \right) \right],
\end{equation}
where $ \nu $ is called the kinematic viscosity coefficient. 
By substituting all the above assumptions and definitions into equations (\ref{eq:continuity})-(\ref{eq:energy}),
we obtain following partial differential equations (PDE) in spherical coordinates. Hence, the continuity equation 
is reduced to the following form

\begin{equation}\label{pde_continuity}
  \frac{1}{r^{2}} \frac{\partial}{\partial r} \left( r^{2} \rho v_{r} \right) + \frac{1}{r \sin \theta} \frac{\partial}{\partial \theta} \left( \rho v_{\theta} \sin\theta \right) = 0.
\end{equation}
The three components of the momentum equation are as
 
\begin{multline}\label{pde_mom1}
\rho \left[ v_{r} \frac{\partial v_{r}}{\partial r}  + \frac{v_{\theta}}{r} \left( \frac{\partial v_{r}}{\partial \theta} - v_{\theta} \right) - \frac{v_{\phi}^{2}}{r} \right] = \\
- \frac{GM \rho}{r^{2}} - \frac{\partial p_\mathrm{gas}}{\partial r}  + \frac{1}{r \sin \theta} \frac{\partial}{\partial \theta} \left( \sin \theta \sigma_{r \theta} \right),
\end{multline}

\begin{multline}\label{mom2}
 \rho \left[  v_{r} \frac{\partial v_{\theta}}{\partial r} + \frac{v_{\theta}}{r} \left( \frac{\partial v_{\theta}}{\partial \theta} + v_{r} \right) - \frac{v_{\phi}^{2}}{r} \cot \theta \right] = \\
- \frac{1}{r} \frac{\partial p_\mathrm{gas}}{\partial \theta} + \frac{1}{r^{3}} \frac{\partial}{\partial r} \left( r^{3} \sigma_{r \theta} \right),
\end{multline}
\begin{multline}\label{mom3}
   \rho \left[ v_{r} \frac{\partial v_{\phi}}{\partial r} + \frac{v_{\theta}}{r} \frac{\partial v_{\phi}}{\partial \theta} + \frac{v_{\phi}}{r} \left( v_{r} +  v_{\theta} \cot\theta \right)  \right] =  \\
 \frac{1}{r^{2}} \frac{\partial}{\partial r} \left( r^{2} \sigma_{r \phi} \right) + \frac{1}{r \sin \theta} \frac{\partial}{\partial \theta} \left( \sin \theta \sigma_{\theta \phi} \right) \\
+ \frac{1}{r} \left( \sigma_{r \phi} + \sigma_{\theta \phi} \cot \theta \right).
\end{multline}

The energy equation is written as

\begin{multline}\label{pde_energy}
	\rho \left[ v_{r} \frac{\partial e}{\partial r}  + \frac{v_{\theta}}{r}  \frac{\partial e}{\partial \theta} \right] - \frac{p_\mathrm{gas}}{\rho} \left[ v_{r} \frac{\partial \rho}{\partial r}   + \frac{v_{\theta}}{r}  \frac{\partial \rho}{\partial \theta} \right]  = \\
	\frac{\partial v_{\theta}}{\partial r} \sigma_{r \theta} + \frac{\partial v_{\phi}}{\partial r} \sigma_{r \phi} + \frac{1}{r} \left( \frac{\partial v_{r}}{\partial \theta} - v_{\theta} \right) \sigma_{r \theta} \\
	+ \frac{1}{r} \frac{\partial v_{\phi}}{\partial \theta} \sigma_{\theta \phi} - \frac{v_{\phi}}{r} \left( \sigma_{r \phi} + \sigma_{\theta \phi} \cot \theta \right) \\
	+ \frac{1}{r^2} \frac{\partial}{\partial r} \left( r^2 \lambda \frac{\partial T}{\partial r} \right) + \frac{1}{r \sin\theta}\frac{\partial}{\partial \theta} \left( \frac{\lambda \sin\theta}{r} \frac{\partial T}{\partial \theta} \right).
\end{multline}

Since we consider all three components of the velocity and the thermal conduction in this study, the energy equation differs from previous 
self-similar solutions such as \citealt{Narayan and Yi 1995} (see, \citealt{Mosallanezhad et al. 2021} for more details). We follow 
\citealt{Tanaka and Menou 2006} and use a standard form of the thermal conduction, where the heat flux 
depends linearly on the local temperature gradient (see also \citealt{Khajenabi and Shadmehri 2013} for more details).  
To preserve self-similarity, the radial dependency of the thermal conductivity coefficient must be
a power-law function of the radius as $ \lambda(r) = \lambda_{0} r^{1/2 - n} $, where $ n $ is
the density index which will be defined in the self-similar solutions (see equation (\ref{rho_selfsimilar})). 
As we explained in Introduction, the key goal of this study is to check the dependency of the physical properties 
of the inflow and wind to the kinematic viscosity coefficient. Thus, based on the numerical simulations as well as analytical 
solutions, we choose three different models of the kinematic viscosity coefficient as

\begin{equation} \label{ST_prescription}
	\nu = \alpha \sqrt{GM r} 
\end{equation}
\begin{equation}\label{SS_prescription}
\nu =  \alpha \left( c_\mathrm{s}^{2} / \Omega_{\!_{K}} \right),
\end{equation}
\begin{equation}\label{ZE_prescription}
\nu = \alpha \left( c_\mathrm{s}^{2} / \Omega_{\!_{K}} \right) f(\theta),
\end{equation}
where $ \Omega_{\!_{K}} (\equiv GM/r^3)^{1/2} $ is the Keplerian angular velocity, 
$c_\mathrm{s}$ is the sound speed, and $ \alpha $ is the viscosity parameter 
related to `alpha' prescription (See \citealt{Shakura and Sunyaev 1973} for more details). 
In all above definitions, the kinematic viscosity coefficient is proportional to $ \nu(r,\theta) \propto r^{1/2} $. 
Note here that equation (\ref{ST_prescription}) is similar to the one chosen by \citealt{Stone et al. 1999} (Model K) and 
also \citealt{Yuan et al. 2012a}; equation(\ref{SS_prescription}) is similar to that chosen by 
\citealt{Shakura and Sunyaev 1973} for thin disk model. 
Also, equation (\ref{ZE_prescription}) is proposed to mimic the effect of MRI 
associated with the Maxwell stress.
We model $ f $ parameter as $ f(\theta) = 1/2 \left( 1 +  | \cos \theta | \right) $,
where $ 1/2 $ is normalization factor. We choose this form because
in a real accretion disk there is no viscosity and the angular momentum is 
transferred by the Maxwell stress associated with the MHD turbulence driven 
by the magnetorotational instability (MRI). The magnetic stress is provided by 
the magnetic field. Since the magnetic field is smooth and coherent with the 
disk height, it can connect to the infinity and would not be zero above the disk. 
In addition, The average stress tensor can be normalized to the gas pressure 
as $ T_{r \phi} = \alpha p_\mathrm{gas} $.  Moreover, the gas pressure gradient 
decreases toward the rotation axis so, $ \alpha $ should increase from 
the equatorial plane to the rotation axis (see, e.g., figures 3 and 6 of \citealt{Bai and Stone 2013}).
We call these three prescriptions of the kinematic viscosity coefficient presented in 
equations (\ref{ST_prescription})-(\ref{ZE_prescription})
`ST', `SS', and `ZE' models, respectively.
In ST model, the viscosity coefficient changes only in the radial direction and its value is constant in $ \theta $ direction.
Whilst, in SS model the viscosity coefficient is proportional to the sound speed squared or equivalently
the temperature of the accretion flow which increases from the equatorial plane toward the rotation axis. Here, we introduce 
ZE model that is more realistic and physical in some senses. In fact, this particular form represents the main features of
accretion flow. In reality, the angular momentum of the accretion disk is transferred by the magnetic stress 
associated with the MHD turbulence driven by MRI. 
In this HD study, we use the shear stress to mimic the magnetic stress in the dynamics 
of the disk. 
In our simple form, i.e., $ \nu \propto \alpha \left( 1 +  | \cos \theta | \right) $, 
the kinematic viscosity coefficient is minimum at the equatorial plane and reaches to 
its maximum at the rotation axis.\footnote{Note that there must be several different 
forms for the kinematic viscosity coefficient which represent similar behavior but 
here we adopt the simplest one to mimic its dependency on $ \theta $ angle. 
This behavior is consistent with \citealt{Bai and Stone 2013} where they 
studied the accretion disk with strong vertical magnetic field and investigated 
MRI to transport angular momentum from the accretion disk.}

To solve the system of the equations in the vertical direction, we adopt the self-similar solutions 
in the radial direction and present a fiducial radius $ r_{\!_{0}} $ as a power-law form of 
$ (r/r_{\!_{0}}) $. The solutions are introduced as

\begin{equation}\label{rho_selfsimilar}
  \rho \left(r, \theta \right) = \rho_{\!_{0}} \left( \frac{r}{r_{\!_{0}}} \right)^{-n} \tilde{\rho}(\theta) ,
\end{equation}

\begin{equation}\label{vr_selfsimilar}
  v_{r} (r, \theta) = v_{\!_{0}} \left( \frac{r}{r_{\!_{0}}} \right)^{-1/2} \tilde{v}_{r}(\theta), 
\end{equation}

\begin{equation}\label{vtheta_selfsimilar}
  v_{\theta} (r, \theta) = v_{\!_{0}} \left( \frac{r}{r_{\!_{0}}} \right)^{-1/2} \tilde{v}_{\theta}(\theta),
\end{equation}

\begin{equation}\label{vphi_selfsimilar}
  v_{\phi} (r, \theta) = v_{\!_{0}} \left( \frac{r}{r_{\!_{0}}} \right)^{-1/2} \widetilde{\Omega}(\theta) \sin(\theta),
\end{equation}

\begin{equation}\label{p_gas_selfsimilar}
  c_\mathrm{s} (r, \theta) = v_{\!_{0}} \left(  \frac{r}{r_{\!_{0}}} \right)^{-1/2} \tilde{c}_\mathrm{s} (\theta),
\end{equation}
where $ r_{\!_{0}} $, $ \rho_{\!_{0}} $, and $ v_{\!_{0}} \left[= \sqrt{GM/r_{\!_{0}}} \right]$, 
are the units of length, density, and  velocity, respectively. The ordinary differential equations (ODEs) can be derived 
by substituting the above self-similar solutions into the PDEs (\ref{pde_continuity})–(\ref{pde_energy}) as

\begin{equation} \label{ode_continuity}
	\left[ \left(\frac{3}{2} - n\right) \tilde{v}_{r} + \frac{\mathrm{d}\tilde{v}_{\theta}}{\mathrm{d}\theta} + \tilde{v}_{\theta} \cot \theta \right] \tilde{\rho} + \tilde{v}_{\theta} \frac{\mathrm{d} \tilde{\rho}}{\mathrm{d}\theta} = 0
\end{equation}

\begin{multline} \label{ode_mom1}
	\tilde{\rho} \left[ - \frac{1}{2} \tilde{v}_r^2 + \tilde{v}_{\theta}\frac{\mathrm{d}\tilde{v}_r}{\mathrm{d}\theta} - \tilde{v}_{\theta}^2 - \widetilde{\Omega}^2 \sin^2 \theta \right]
	= - \rho  \\
	+ \left( n + 1 \right) p_\mathrm{g} 
	+ \tilde{\rho} \tilde{\nu} \left( \frac{\mathrm{d}^{2}  \tilde{v}_{r}}{\mathrm{d} \theta^{2}}  - \frac{3}{2} \frac{\mathrm{d}  \tilde{v}_{\theta}}{\mathrm{d} \theta} \right) \\
	+ \left( \frac{\mathrm{d}  \tilde{v}_{r}}{\mathrm{d} \theta}  - \frac{3}{2}  \tilde{v}_{\theta}  \right) \left[  \frac{\mathrm{d} \tilde{\nu}}{\mathrm{d} \theta}  \tilde{\rho}  +  \tilde{\nu} \frac{\mathrm{d}  \tilde{\rho}}{\mathrm{d} \theta} +  \tilde{\rho}  \tilde{\nu} \cot \theta  \right] ,
\end{multline}

\begin{multline} \label{ode_mom2}
	\tilde{\rho} \left[ \frac{1}{2} \tilde{v}_{r} \tilde{v}_{\theta} + \tilde{v}_{\theta} \frac{\mathrm{d} \tilde{v}_{\theta}}{\mathrm{d} \theta} -\widetilde{\Omega}^2 \sin \theta \cos \theta \right] = - \frac{\mathrm{d} p_\mathrm{g}}{\mathrm{d} \theta} \\
	-\left( n - 2 \right)  \tilde{\rho} \tilde{\nu}  \left[ \frac{\mathrm{d} \tilde{v}_{r}}{\mathrm{d} \theta}  - \frac{3}{2} \tilde{v}_{\theta}  \right],
\end{multline}

\begin{multline}\label{ode_mom3}
\tilde{\rho} \left[ \frac{1}{2} \tilde{v}_{r} \widetilde{\Omega} + 2 \tilde{v}_{\theta} \widetilde{\Omega} \cot \theta + \tilde{v}_{\theta} \frac{\mathrm{d}\widetilde{\Omega}}{\mathrm{d}\theta} \right] =  \frac{\mathrm{d} \widetilde{\Omega}}{\mathrm{d} \theta} \left[ \tilde{\nu} \frac{\mathrm{d} \tilde{\rho}}{\mathrm{d} \theta} + \tilde{\rho} \frac{\mathrm{d} \tilde{\nu}}{\mathrm{d}\theta} \right] \\
+ \tilde{\rho} \tilde{\nu} \left[ \frac{3}{2} \left(n - 2 \right) \widetilde{\Omega} +  3 \frac{\mathrm{d} \widetilde{\Omega}}{\mathrm{d} \theta} \cot \theta + \frac{\mathrm{d}^{2} \widetilde{\Omega}} {\mathrm{d} \theta^{2}} \right]
\end{multline}

\begin{multline}\label{ode_energy}
\left( n  - \frac{1}{ \gamma - 1} \right) \tilde{p}_\mathrm{g} \tilde{v}_{r} + \frac{\tilde{v}_{\theta}}{\gamma - 1}  \left( \frac{\mathrm{d} \tilde{p}_\mathrm{g}}{\mathrm{d} \theta} - \gamma \frac{\tilde{p}_\mathrm{g}}{\tilde{\rho}} \frac{\mathrm{d} \tilde{\rho}}{\mathrm{d} \theta} \right) = \\ 
\tilde{\rho} \tilde{\nu} \Biggr\{ \left( \frac{\mathrm{d} \tilde{v}_{r}}{\mathrm{d} \theta} - \frac{3}{2} \tilde{v}_{\theta} \right)^{2} + \left[ \frac{9}{4} \widetilde{\Omega}^{2} + \left( \frac{\mathrm{d} \widetilde{\Omega}}{\mathrm{d}\theta} \right)^{2} \right] \sin^{2} \theta   \Biggr\} \\
+ \lambda_{0} \left[ \left( n - \frac{1}{2} \right) \tilde{c}_{s}^{2} + \frac{1}{\sin\theta} \frac{\mathrm{d}}{\mathrm{d} \theta} \left(\sin \theta \frac{\mathrm{d} \tilde{c}_{s}^{2}}{\mathrm{d} \theta} \right)  \right].
\end{multline}

In the above system of equations, $ \tilde{p}_\mathrm{g}(\theta) $ is the dimensionless form of the gas pressure defined as \footnote{For simplicity, we remove the $ \theta $ dependency of the variables in equations (\ref{ode_continuity})-(\ref{ode_energy})}

\begin{equation}
   \tilde{p}_\mathrm{g}(\theta) = \tilde{\rho}(\theta) \tilde{c}_{s}^{2} (\theta).
\end{equation}

The dimensionless forms of the kinematic viscosity coefficient for three models will be also written as

\begin{equation}
	\tilde{\nu} (\theta) = \alpha, 
\end{equation}
\begin{equation}
	\tilde{\nu} (\theta) = \alpha \tilde{c}_{s}^{2}(\theta), 
\end{equation}
\begin{equation}
	\tilde{\nu} (\theta) = \frac{1}{2} \alpha \tilde{c}_{s}^{2}(\theta) \left( 1 + | \cos \theta | \right). 
\end{equation}

\begin{figure*}[ht!]
\includegraphics[width=0.5\textwidth]{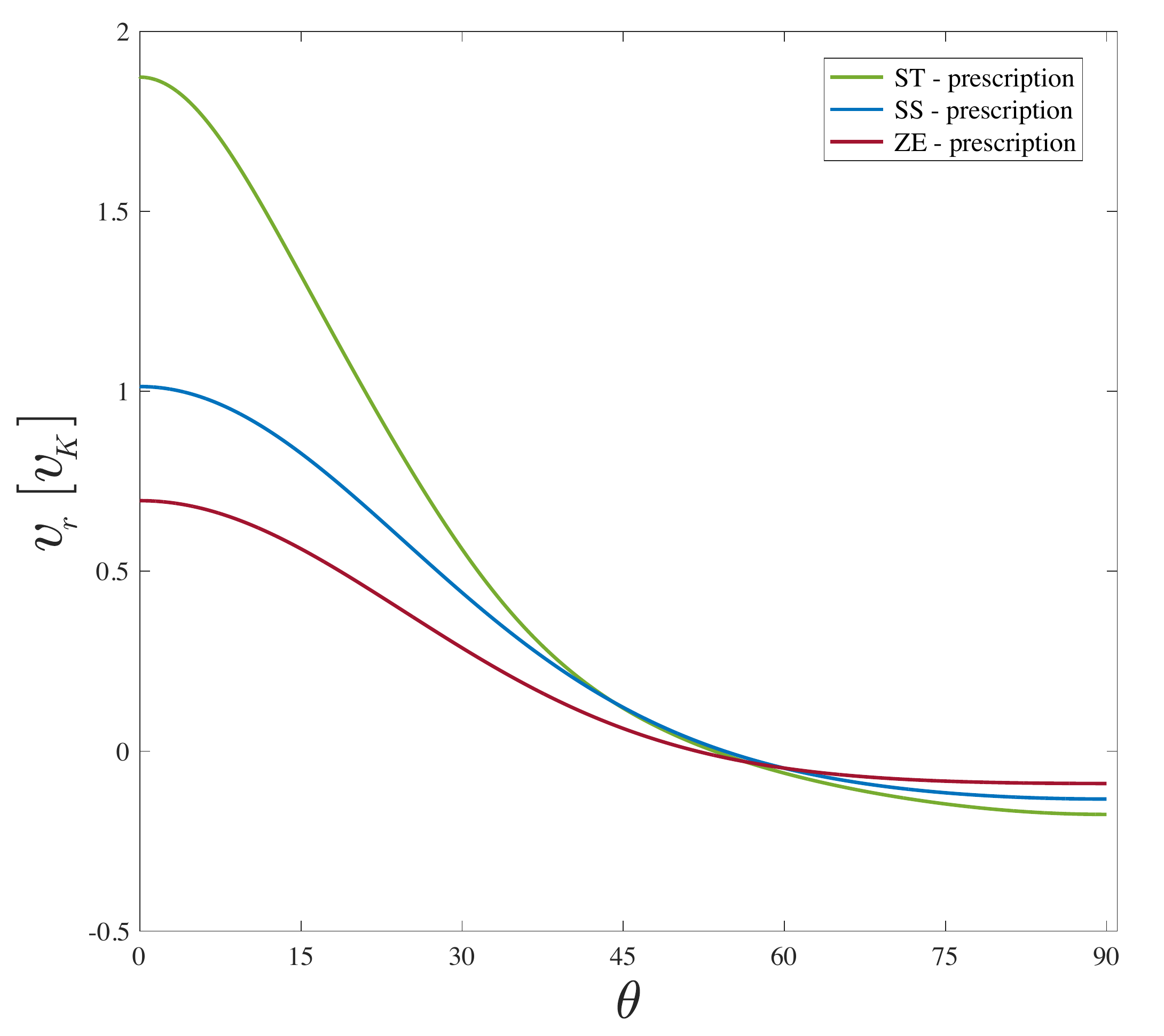}
\centering
\caption{Latitudinal profile of radial velocity in the unit of Keplerian velocity, $ v_{\!_{K}} $. 
The green, blue, and red curves are corresponding to three models of the kinematic viscosity coefficient, ST; $ \nu = \alpha \sqrt{GM r} $,
SS; $ \nu =  \alpha \left( c_\mathrm{s}^{2} / \Omega_{\!_{K}} \right), $ and ZE; $ \nu =  \frac{1}{2}  \alpha \left( c_\mathrm{s}^{2} / \Omega_{\!_{K}} \right) \left( 1 +  | \cos \theta | \right) $. The inflow region is in the range of $ 45^{\circ} < \theta \le 90^{\circ} $ for all models. Here, $ \alpha = 0.1 $, $ n = 0.5 $, $ \gamma = 5/3 $, and $ \lambda_{0} = 0.02 $.    
\label{vx1}}
\end{figure*}

The above ODEs, i.e., equations (\ref{ode_continuity})-(\ref{ode_energy}) consist of five 
physical variables: $ \tilde{v}_{r} (\theta) $, $ \tilde{v}_{\theta} (\theta) $, $ \widetilde{\Omega} (\theta) $, $ \tilde{\rho}(\theta) $, and $ \tilde{c}_\mathrm{s}(\theta) $. Following \citealt{Zeraatgari et al. 2020} and \citealt{Mosallanezhad et al. 2021}, the computational domain will be extended from equatorial plane, $ \theta = \pi/2 $, to the rotation axis, $ \theta = 0 $. In addition, all physical variables are assumed to be even symmetric, continuous, and differentiable at both boundaries. We also include the latitudinal component of the velocity, $ \tilde{v}_{\theta} $, with zero values at both equatorial plane and the rotation axis. The following boundary conditions will be imposed at $ \theta = \pi/2 $ and $ \theta = 0 $: 

\begin{equation} \label{boundary_conditions}
	\frac{\mathrm{d} \tilde{\rho}}{ \mathrm{d} \theta} = \frac{\mathrm{d} \tilde{c}_\mathrm{s}}{\mathrm{d} \theta} 
	=  \frac{\mathrm{d} \widetilde{\Omega}}{\mathrm{d} \theta} = \frac{\mathrm{d} \tilde{v}_{r}}{ \mathrm{d} \theta} =  \tilde{v}_{\theta} = 0.
\end{equation}

To integrate the ODEs, the relaxation method will be adopted 
with the resolution of 5000 stretch grids as follows: 
from $ \theta = \pi/2 $ to $ \theta = \pi/4 $ the grid size ratio is set as 
$ \mathrm{d}\theta_\mathrm{i+1}/ \mathrm{d}\theta_\mathrm{i} = 1.003$,
while from $ \theta = \pi/4 $ to $ \theta = 0 $ the grid size ratio is set as 
$ \mathrm{d}\theta_\mathrm{i+1}/ \mathrm{d}\theta_\mathrm{i} =  0.997 $
The absolute error tolerance is considered to be $ 10^{-15} $. For the appropriate initial guess,
we use \textit{Fourier cosine series} for all physical variables except $ \tilde{v}
_{\theta} $ 
where we use \textit{Fourier sine series}. Our solutions will satisfy the boundary conditions 
at both ends which ensures to have well-behaved solutions in the whole computational domain.
In the next section, we will explain in detail the behaviors of the physical variables for three different viscosity 
models.

\begin{figure*}[ht!]
\includegraphics[width=\textwidth]{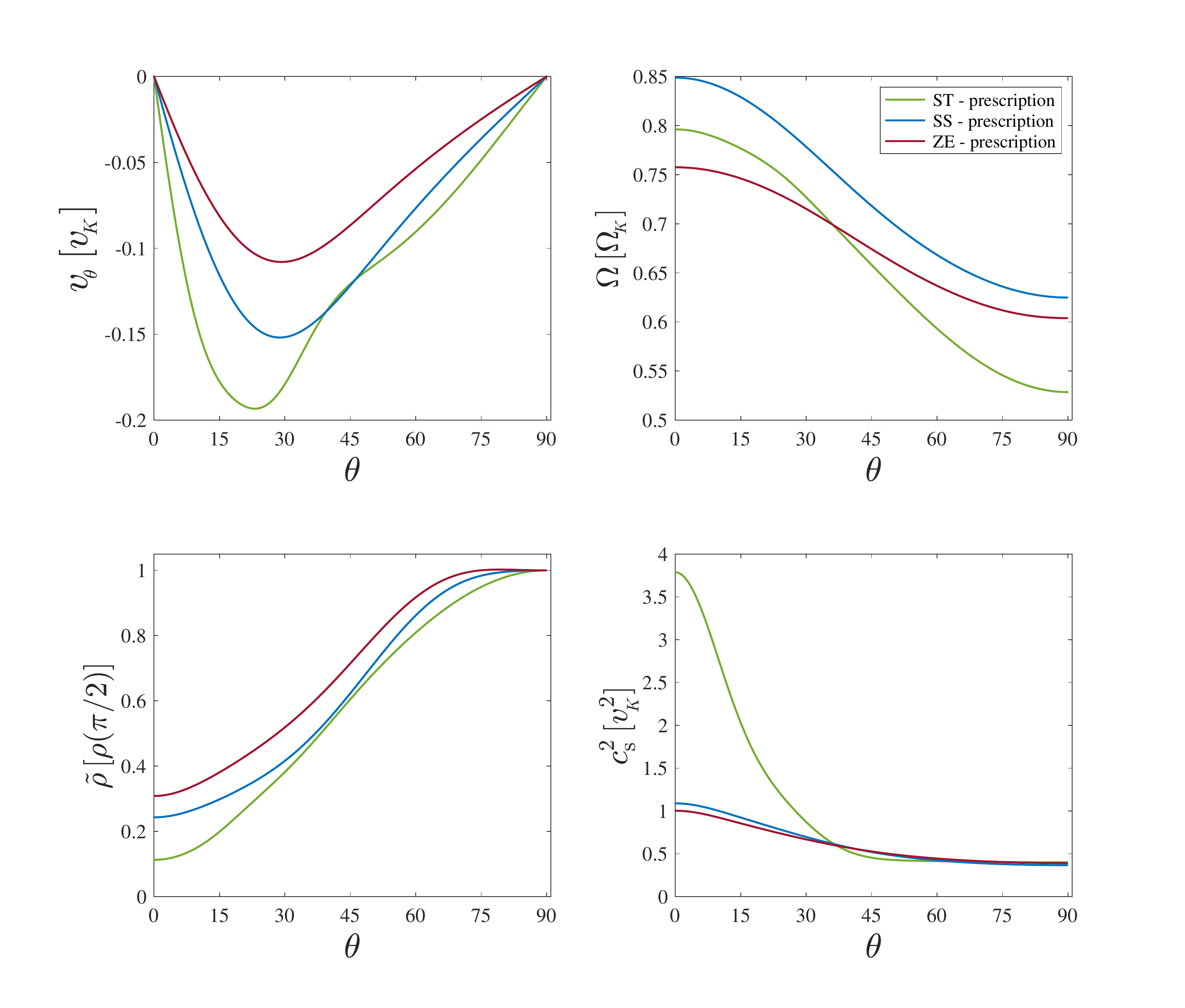}
\centering
\caption{Latitudinal profiles of the physical variables. \textit{Top left:} latitudinal velocity, $ v_{\theta} $, in the unit of Keplerian velocity, $ v_{\!_{K}} $.
\textit{Top right:} angular velocity, $ \Omega $, in the unit of Keplerian angular velocity, $ \Omega_{\!_{K}} $.
\textit{Bottom left:} density in the unit of the density at the equatorial plane.
\textit{Bottom right:} sound speed squared, $ c_s^2 $, in the unit of $ v_{\!_{K}}^{2} $. 
The green, blue, and red curves are corresponding to three models of the kinematic viscosity prescription, `ST', `SS', and `ZE', respectively. Here, $ \alpha = 0.1 $, $ n = 0.5 $, $ \gamma = 5/3 $, and $ \lambda_{0} = 0.02 $.    
\label{variables}}
\end{figure*}

\begin{deluxetable*}{cccccc}
\tablenum{1}
\tablecaption{Physical properties of the inflow and wind\label{tab:properties}}
\tablewidth{0pt}
\tablehead{
\colhead{Model} & \colhead{Kinematic viscosity coefficient} & \colhead{$ L_\mathrm{inflow} $} & \colhead{$ L_\mathrm{wind} $} &
\colhead{Bernoulli parameter} & \colhead{$ v_{p} $} \\
\colhead{} & \colhead{$ \nu $} & \colhead{$ (L_\mathrm{K}) $} & \colhead{$ (L_\mathrm{K}) $} &
\colhead{$ (v_\mathrm{\!_{K}}^{2}) $} & \colhead{ $ (v_\mathrm{\!_{K}}) $}
}
\startdata
ST & $ \nu = \alpha \sqrt{GM r} $ & 0.54 & 0.73  & 1.24 & 0.75  \\
SS & $ \nu =  \alpha \left( c_\mathrm{s}^{2} / \Omega_{\!_{K}} \right) $ & 0.63  & 0.78  & 0.34 & 0.52  \\
ZE & $ \nu =  \frac{1}{2}  \alpha \left( c_\mathrm{s}^{2} / \Omega_{\!_{K}} \right) \left( 1 +  | \cos \theta | \right) $ & 0.61  & 0.72 & 0.22 & 0.37  \\
\enddata
\end{deluxetable*}

\section{Numerical Results} \label{sec:results}
The system of ODEs are solved numerically by using relaxation method
which is utilized in two-point boundary value problems. The equations 
(\ref{ode_continuity})-(\ref{ode_energy}) are integrated from the equatorial 
plane ($ \theta = \pi/2 $), where the maximum density is located, to the 
rotation axis ($ \theta = 0 $). Our numerical method might guarantee 
that the solutions would be mathematically self-consistent in $ r-\theta $ space.
The parameters we adopt are $ \alpha = 0.1 $, $ n = 0.5 $, $ \gamma = 5/3 $, 
and $ \lambda_{0} = 0.02 $. We plotted the latitudinal profiles of the physical 
variables in Figures \ref{vx1} and \ref{variables}. The latitudinal profile of $ v_r $ 
in Figure \ref{vx1} is plotted for three prescriptions of the kinematic viscosity 
coefficient, i.e., $ \nu = \alpha \sqrt{GM r} $ (ST), $ \nu =  \alpha \left( c_\mathrm{s}^{2} 
/ \Omega_{\!_{K}} \right) $ (SS), and $ \nu =  \frac{1}{2}  \alpha \left( c_\mathrm{s}^{2} 
/ \Omega_{\!_{K}} \right) \left( 1 +  | \cos \theta | \right) $ (ZE) in green, blue, 
and red colors, respectively. The radial velocity is negative in the inflow 
region extended in the range of $ 45^{\circ} < \theta \le 90^{\circ} $ for all models. 
The ST prescription shows a higher radial velocity in the inflow and wind region 
and the wind velocity becomes around twice the Keplerian velocity at the rotation axis.
As we mentioned earlier, the ST prescription is similar to the one presented in 
\citealt{Stone et al. 1999} (model K) and \citealt{Yuan et al. 2012b} which 
the kinematic viscosity coefficient is only proportional to the radius, therefore 
it is constant in the vertical direction. As this figure shows, SS and ZE models 
in comparison with ST model have smaller radial velocity in wind region and 
it is clear that the smallest radial velocity is for ZE model around $ \approx 0.7 v_{\!_{K}} $.
Although we did not include magnetic field in our study, we mimic the effect of MRI 
by considering different prescriptions of kinematic viscosity coefficient.
Therefore, this behavior of our model comes from this fact that the entire 
disk becomes magnetically dominated where the strong magnetic pressure 
gradient would suppress the radial velocity of the wind in the region far 
from the central black hole.

\begin{figure*}[ht!]
\includegraphics[width=0.85\textwidth]{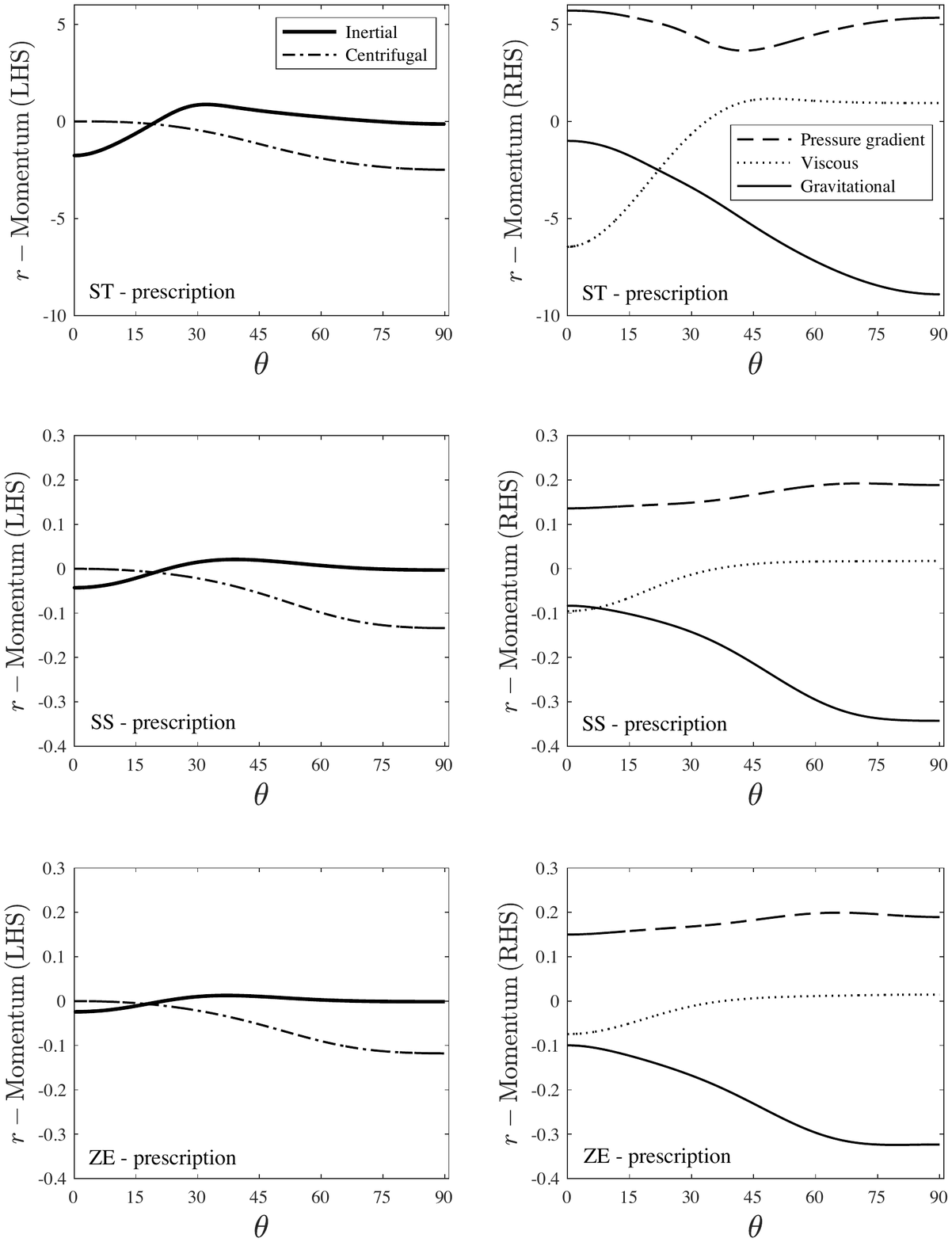}
\centering
\caption{Latitudinal profiles of the terms in the $ r $-component of the momentum equation for three models of 
kinematic viscosity prescription, ST (top panels), SS (middle panels), and ZE (bottom panels) which
left and right panels are corresponding to the left- and right-hand sides of this equation (eq. [\ref{ode_mom1}]).  
\label{mom_r}}
\end{figure*}

In Figure \ref{variables}, the latitudinal profile of the four physical variables,
latitudinal velocity, $ v_{\theta}$, in the unit of the Keplerian velocity, $ v_{\!_{K}} $ (top left panel), 
the angular velocity, $ \Omega $ in the unit of the Keplerian angular velocity (top right panel),
density in the unit of density at the equatorial plane (bottom left panel),
and sound speed squared in the unit of $ v_{\!_{K}}^{2} $ (bottom right panel) are plotted.
As you can see, for all models, $ v_{\theta} $ is always negative in the range of 
$ -0.2 < v_{\theta} < -0.1 $ and has a minimum located at
$ 15^{\circ} < \theta  < 30^{\circ} $.
It is zero at $ \theta = 0 $ and $ \theta = \pi/2 $ as boundary conditions.
The ST model has the minimum value of $ v_{\theta} $ happened 
around $ -0.2 $ at $ \theta \sim 24^{\circ} $.
We note that as in our solution the density index is $ 0 < n < 3/2 $,
the net mass accretion rate $ [\dot{M} = 2\pi \int _{0}^{\pi} \rho (r, \theta) v_{r} (r, \theta) r^{2} \sin \theta\, d\theta] $
becomes zero. In essence, the self-similar solutions are applied far away from 
the black hole where the gravity of the black hole is not so strong.
Therefore, at that region, we would find that the net mass accretion 
rate is actually negligible compare to the mass inflow and outflow rates. 
Then, we can model this case with zero mass accretion rate.
To find whether the wind can escape from the system and go to infinity
we quantitatively calculate some properties of the wind summarized 
in Table \ref{tab:properties}.

According to Figure \ref{vx1} and this panel, the poloidal velocities 
($ | \vec{v}_{p} | = \sqrt{ v_{r}^{2} + v_{\theta}^2 } $) in ST and SS models 
become greater than that in ZE model.
We define the mass flux-weighted of the inflow and the wind as

\begin{equation} \label{mass_flux_inflow}
	q_\mathrm{inflow}(r) = \frac{4 \pi r^{2} \int^{\pi/2}_{0} \rho \, q \, \mathrm{min}(v_{r},0) \sin \theta d\theta}
	{4 \pi r^{2} \int^{\pi/2}_{0} \rho \, \mathrm{min}(v_{r},0) \sin \theta d\theta},
\end{equation}

\begin{equation} \label{mass_flux_outflow}
	q_\mathrm{wind}(r) = \frac{4 \pi r^{2} \int^{\pi/2}_{0} \rho q \, \mathrm{max}(v_{r},0) \sin \theta d\theta}
	{4 \pi r^{2} \int^{\pi/2}_{0} \rho \, \mathrm{max}(v_{r},0) \sin \theta d\theta},
\end{equation}
where $ q $ is a physical quantity. The mass flux-weighted 
poloidal velocities of the wind for three models, ST, SS, and ZE 
are brought in the last column of Table \ref{tab:properties}. 
Our model, ZE prescription, presents the minimum poloidal velocity,
$ v_p = 0.37\, v_{\!_K} $, whilst the ST model shows the maximum 
one equal to $ v_p = 0.75 \, v_{\!_K} $. These results are in agreement 
with the numerical simulation of \citealt{Yuan et al. 2015}.

\begin{figure*}[ht!]
\includegraphics[width=0.85\textwidth]{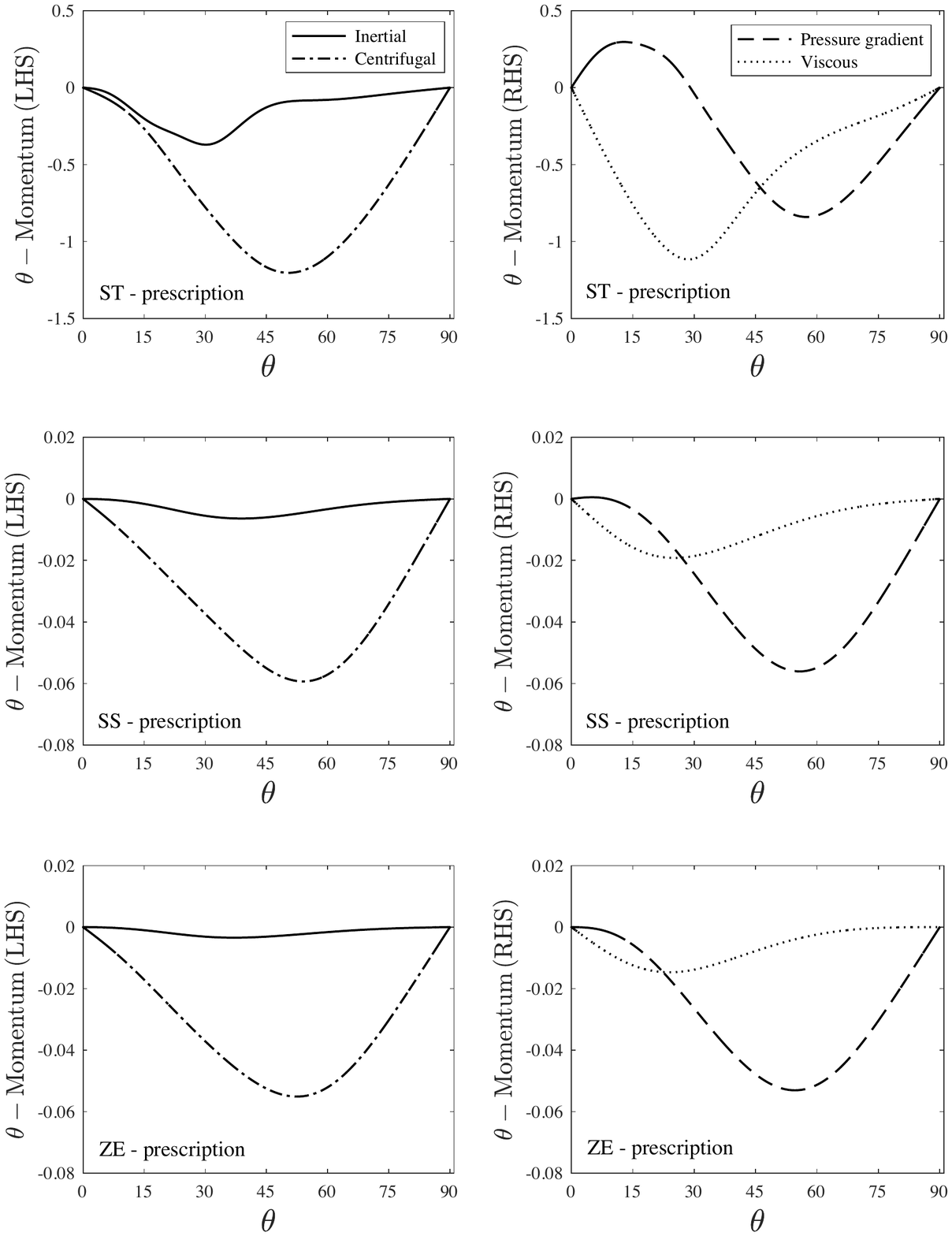}
\centering
\caption{ Latitudinal profiles of terms in the $ \theta $-component of the momentum equation for three models of 
kinematic viscosity prescription, ST (top panels), SS (middle panels), and ZE (bottom panels) which
left and right panels are corresponding to left- and right-hand sides of this equation (eq. [\ref{ode_mom2}]).}
\label{mom_theta}
\end{figure*}

In the top right panel of Figure \ref{variables}, it is shown the behavior 
of the angular velocity, $ \Omega $, in the latitudinal direction.
As it is clear, the angular velocity in all models is sub-Keplerian.
From the equatorial plane to the rotation axis, it has increasing trend.
Since the maximum angular velocity, for all models, occurs in the polar region, it is 
strongly suggested that the wind can vertically transport the angular momentum to the out of the system.
As it was noted before, we propose ZE model for kinematic viscosity coefficient
to modify the behavior of MRI. Therefore, we expect the entire disk becomes 
magnetically dominated and strong manetic field would lead to substantial sub-Keplerian rotation
for our ZE model.
We also calculated the mass flux-weighted angular momentum of 
the inflow and wind in the unit of the Keplerian angular momentum,
so the results are given in the third and fourth columns of Table \ref{tab:properties}, respectively.
As it can be seen, in all models the mass flux-weighted angular momentum of the wind are larger than 
those for the inflow. As an example, the mass flux-weighted angular momentum of 
the inflow and wind for SS model are calculated as

\begin{equation}
L_\mathrm{inflow} =0.63 L_{\mathrm{K}},
\end{equation}
\begin{equation}
L_\mathrm{wind} =0.78 L_{\mathrm{K}},
\end{equation}
which comparing with two other models we find that SS model is 
the most efficient prescription to extract the angular momentum 
from the system.

In the bottom left panel of Figure \ref{variables}, the trend of 
the density is plotted. To have a similar density scale as 
\citealt{Xu and Chen 1997} and \citealt{Xue and Wang 2005}, we 
normalized the density to the maximum density at the equatorial 
plane. In all models, the density decreases from the disk mid-plane 
to the rotation axis. More precisely, based on the density curves, 
the densities reach to (0.1-0.3) of their maximum values at the rotation axis.
Note that based on our self-similar solutions, the mass 
density profile increases about one order of magnitude 
which is consistent with the previous self-similar studies 
such as \citealt{Narayan and Yi 1995} (figure 1) and 
\citealt{Tanaka and Menou 2006} (figure 3). However, 
the numerical simulations 
of hot accretion flow (see, e.g., figures 1 and 5 of 
\citealt{Yuan and Bu 2010}) show that the mass density 
close to the polar axis is lower than that on the equatorial plane 
by several (two or three, or even larger) orders of magnitude.

The latitudinal profile of the sound speed, $ c_s^2 $, is shown 
in the bottom right panel of Figure \ref{variables}. This plot 
shows that the sound speed in both SS and ZE models are almost 
in the same range and small, while it is much higher in the ST model 
in the wind region. 
It shows that in ST model the flow is easily heated to a temperature 
about the virial. This is in good agreement with MHD simulations 
of \citealt{Yuan et al. 2012a}.

\begin{figure*}[ht!]
\includegraphics[width=\textwidth]{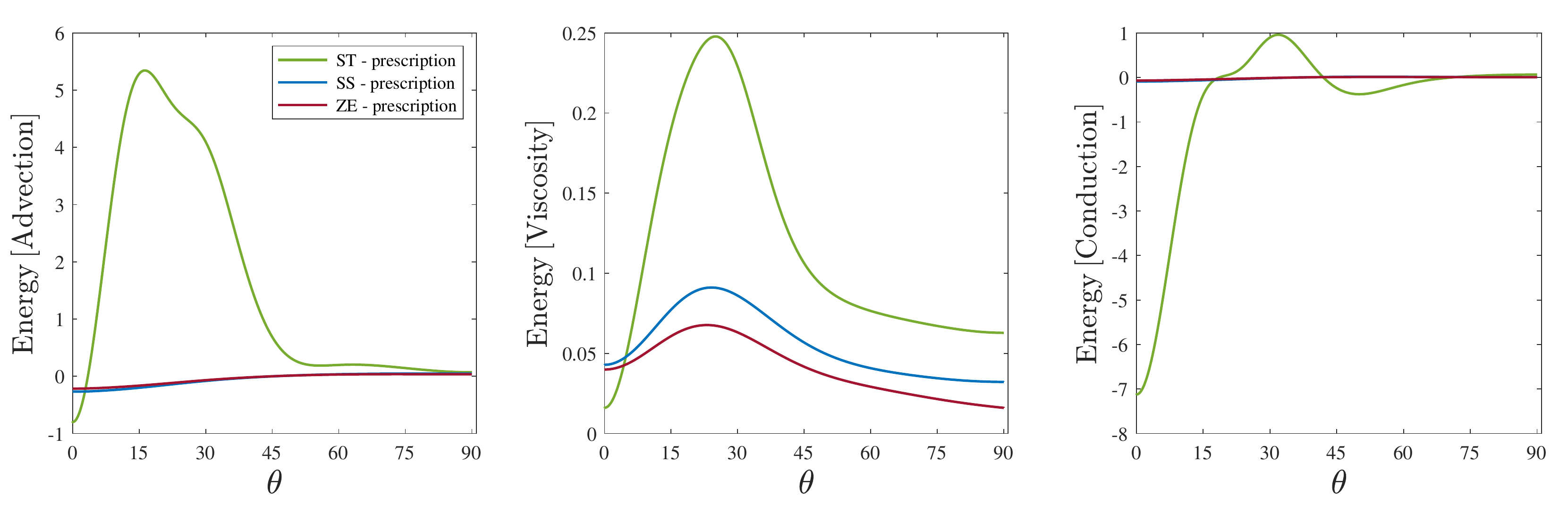}
\centering
\caption{Latitudinal profiles of the advection (left panel), viscosity (middle panel), and conduction (right panel)
terms in the energy equation for three different models of kinematic viscosity prescription, ST (green), SS (blue), ZE (red).}
\label{energy}
\end{figure*}

To calculate the wind properties, we define the Bernoulli parameter as

\begin{equation}
Be(r) = \frac{v^2}{2} + \frac{\gamma p}{(\gamma - 1)\rho} - \frac{GM}{r},
\end{equation}
where on the right-hand side of this equation, the first, second, and the third terms are
corresponding to the kinetic energy, enthalpy, and gravitational energy, respectively.
The mass flux-weighted Bernoulli parameter of the wind are given in
the fifth column of Table \ref{tab:properties} for three models of kinematic viscosity prescription.
For all models, the Bernoulli parameter of the wind is positive which shows the flow 
has enough energy to overcome the gravity of the central object and take away from the system. 
As we mentioned earlier, the density index is set to $ n =0.5 $ in 
our current study, which causes the net mass accretion rate becomes zero.
At the region far away from the black hole, the wind can be produced at 
any radius based on our self-similar solution. For instance, when the gas 
approaches to a given radius, i.e., $ r = r_{0} $, the release of the gravitational 
energy is enough to produce the wind at that radius. However, our self-similar 
model cannot be applied to observations of high velocity outflow (such as ultra-fast 
outflows, UFOs) arguably produced very close to the black hole.

To investigate the momentum balance in different models of kinematic viscosity prescription, 
we plotted latitudinal profiles of various terms in the radial and vertical components 
of the momentum equation (equations (\ref{ode_mom1}) and (\ref{ode_mom2})) in 
Figures \ref{mom_r} and \ref{mom_theta}, respectively. 
In the left panels of Figure \ref{mom_r}, radial components of the inertial (thick solid lines)
and the centrifugal (dash-dotted lines) terms are plotted for three models of kinematic viscosity 
prescriptions, ST, SS, and ZE based on radial component of the momentum equation (see LHS of equation (\ref{ode_mom1})).
While in the right panels of this figure, the gravitational (thin solid lines), 
pressure gradient (dashed lines), and viscous (dotted lines)
terms are calculated from the right-hand side of the r-momentum.
As you can see, in the left panels of this figure, the absolute value of the radial centrifugal term is 
larger than the inertial term at the equatorial plane in all models.
The inertial term becomes significant in the polar region and the maximum 
amount is for ST model.
From the right panels it is clear that the gravitational term plays the important role 
in the equatorial plane while it drops in the polar region where the density reaches to its minimum.
The pressure gradient and viscous terms become substantial in the wind region for all models. 
Comparing three models, the maximal amount of viscous term is shown in ST model.
The amount and behavior of the pressure gradient is nearly the same
in SS and ZE models, it decreases from the equatorial plane toward the rotation axis, 
while it is high at both the equatorial plane and the rotation axis for ST model. 
Further, the amount of the pressure gradient term in ST model should also be the highest one which 
could balance the high amount of the viscous term in this model at the rotation axis.

The latitudinal profiles of the left- and right-hand side terms in $\theta$-momentum
are shown in Figure \ref{mom_theta} which is similar to Figure \ref{mom_r}. 
This figure also shows a comparison between three different prescriptions of the kinematic viscosity coefficient, 
ST (top panels), SS (middle panels), ZE (bottom panels).
As you can see, the vertical component of the centrifugal (dash-dotted lines) 
and inertial (solid lines) terms are shown in the left panels, 
while the pressure gradient (dashed line) 
and viscous (dotted lines) terms are plotted in the right panels. 
From the left panels, it is clear that the vertical component of 
inertial term is not important in none of the models.
Also, in the right panels, viscous term, nowhere of the flow, 
does show a dominant role in SS and ZE models. 
Instead, in ST model, this term is considerable everywhere in 
the flow excluding at the equatorial and rotation axis which it is zero.
This figure points out that in the vertical component of the momentum equation, the latitudinal pressure gradient,
projected centrifugal force and viscosity would balance each other.
Moreover, the most important feature of Figures \ref{mom_r} and \ref{mom_theta}
is that through the momentum balance analysis, one cannot determine 
the inflow and wind regions, so the winds 
can be specified through energetics treatment.

Therefore, to describe the wind, we examined the energy 
balance in three different models of the kinematic viscosity 
prescription. Figure \ref{energy} represents 
the latitudinal profiles of advection (left panel), viscous (middle panel), 
and conduction (right panel) terms in the energy equation.
The energy advection is on the left-hand side of equation (\ref{ode_energy}) 
while the viscosity and the thermal conduction are on the right-hand side.
As you can see, the viscous term is positive from the equatorial plane
toward the rotation axis and heats the flow.
At the equatorial plane, the conduction term is positive 
and is a heat source so the advection heating acts as a cooling term with positive sign.
On the contrary, near the rotation axis the amount of the viscous term
declines so the advection term has to balance the thermal conduction.
As the sign of the conduction term is negative, it is
a cooling term hence the advection works as a heating 
despite being negative. From this results, it is suggested that to launch wind from the system
near the rotation axis there should be significant amount of 
cooling that advection behaves as a heating.
It should be noted that the larger amount of all energy terms in ST model with respect to other models
is only because the scales are different.

\section{Summary and Discussion}\label{sec:summary_discussion}
In this work, we have solved the HD equations of hot accretion flows 
associated with thermal conduction in spherical coordinates $(r,\theta,\phi)$ and in two dimensions
using self-similar solutions in the radial direction. We assumed the flow is 
axisymmetric and in steady state. We also considered three components of the 
viscous stress tensor, i.e., $ r\theta, r\phi $, and $ \theta\phi $.
We examined three different models for kinematic viscosity coefficient,
ST model; the kinematic viscosity coefficient depends only on $ r $, SS model;
the kinematic viscosity coefficient is proportional to the sound speed squared, and
ZE model; the kinematic viscosity coefficient is proportional to $ (1 + |\cos \theta|) $.
In ZE model that we proposed, the kinematic
viscosity coefficient is minimum at the equatorial
plane and reaches to its maximum at the rotation axis
to satisfy MHD numerical simulations which predict  
$ \alpha $ increases with decreasing $ \theta $.
We made use of relaxation method and integrated the ODEs 
from the equatorial plane to the rotation axis. 
Our motivation was to see which model of 
the kinematic viscosity coefficient
could produce strong wind far away from the black hole in the hot accretion flow.
The physical variables of the inflow and wind were obtained
for three models of the kinematic viscosity coefficient.
The positive Bernoulli parameter in all three models can be admittedly interpreted as 
wind can exist in hot accretion flows.
Then, the inflow region for all models were attained in the range of $ 45^{\circ} < \theta \leq 90^{\circ} $.
In addition, our model illustrates that the radial velocity of the wind near the rotation 
axis is lower than two other models mainly due to the existence of strong magnetic 
pressure gradient at the surface of the disk.  
We found that the poloidal velocity in the ST model is higher than that in 
two other models. Additionally, the gas rotates at sub-Keplerian velocities for
all three models. Comparing the angular velocity profile in three models, 
it is clearly demonstrated that in the model we suggested wind is less efficient 
than that in two other models to extract the angular
momentum outward due to the behavior of MRI.
We treated the momentum balance in the radial and polar directions for three models
of the kinematic viscosity coefficient and found the pressure gradient and also viscous terms in both directions in ST model
is stronger than those in two other models. It is imperative that the inflow and wind regions
cannot be distinguished by momentum balance analysis
noting that we need a energetics examination for a detailed description of the winds.
In this regard, we investigated the energy balance in three different prescriptions of the kinematic viscosity coefficient.
Subsequently, our results implied that in inflow region where the viscous term is large enough
the conduction is a heat source so advective acts as a cooling.
While near the rotation axis where the viscosity decreases,
the conduction is a cooling term and advection acts as a heating.
As a Consequence, to produce wind in the high latitudes of the accretion flow 
there should be sufficient cooling that the advection acts as a heating.
The solutions obtained in this paper might be applicable for 
numerical hydrodynamical simulations of hot accretion flow.
Therefore, it is worthwhile to be revisited by the numerical simulations.

\section*{Aknowledgments}
The authors would like to thank the referee for constructive comments and suggestions
which improved the manuscript.
F. Z. Z. is supported by the National Natural Science Foundation of China (grant No. 12003021) 
and also the China Postdoctoral Science Foundation (grant No. 2019M663664). 
L. M. is supported by the Science Challenge Project of China (grant No. TZ2016002). 
A.M. is supported by the China Postdoctoral Science Foundation (grant No. 2020M673371). 
A. M. also acknowledges the support of Dr. X. D. Zhang at the Network Information Center of Xi'an Jiaotong University. 
The computation has made use of the High Performance Computing (HPC) platform of Xi’an Jiaotong University.


%


\bibliography{sample63}{}
\bibliographystyle{aasjournal}



\end{document}